\documentclass[a4paper,12pt, epsfig]{article}
\usepackage{epsfig}
\usepackage{amssymb}
\usepackage{amsfonts}

\newskip\humongous \humongous=0pt plus 1000pt minus 1000pt

\newif\ifdtup

\catcode`\@=11

\@addtoreset{equation}{section}
\def\theequation{\thesection.\arabic{equation}}

\def\@normalsize{\@setsize\normalsize{15pt}\xiipt\@xiipt
\abovedisplayskip 14pt plus3pt minus3pt%
\belowdisplayskip \abovedisplayskip
\abovedisplayshortskip \z@ plus3pt%
\belowdisplayshortskip 7pt plus3.5pt minus0pt}

\def\small{\@setsize\small{13.6pt}\xipt\@xipt
\abovedisplayskip 13pt plus3pt minus3pt%
\belowdisplayskip \abovedisplayskip
\abovedisplayshortskip \z@ plus3pt%
\belowdisplayshortskip 7pt plus3.5pt minus0pt
\def\@listi{\parsep 4.5pt plus 2pt minus 1pt
     \itemsep \parsep
     \topsep 9pt plus 3pt minus 3pt}}

\relax

\catcode`@=12

\catcode`\@=11

\def\section{\@startsection{section}{1}{\z@}{3.5ex plus 1ex minus
   .2ex}{2.3ex plus .2ex}{\large\bf}}

\def\thesection{\arabic{section}}
\def\thesubsection{\arabic{section}.\arabic{subsection}}

\def\appendix{\setcounter{section}{0}
 \def\thesection{Appendix \Alph{section}}
 \def\thesubsection{\Alph{section}.\arabic{subsection}}
 \def\theequation{\Alph{section}.\arabic{equation}}}

\def\SymBoxes#1#2#3#4{\newdimen\un@t \un@t#3%
\raisebox{#1}{\rule{#2\un@t}{#4}\hskip-#2\un@t
\@tempdimb\un@t \advance\@tempdimb by-#4\@tempcntb#2\relax%
\@whilenum{\@tempcntb>0}\do{
\rule{#4}{\un@t}\hskip\@tempdimb \advance\@tempcntb by\m@ne}%
\hskip-#2\un@t \rule[\un@t]{#2\un@t}{#4}%
\rule[\un@t]{#4}{#4}\hskip-#4
\rule{#4}{\un@t}}\hskip-#4}                

\begin{document}

\newcommand{\beq}{\begin{equation}}
\newcommand{\eeq}{\end{equation}}
\newcommand{\bea}{\begin{eqnarray}}
\newcommand{\eea}{\end{eqnarray}}
\newcommand{\beas}{\begin{eqnarray*}}
\newcommand{\eeas}{\end{eqnarray*}}
\newcommand{\defi}{\stackrel{\rm def}{=}}
\newcommand{\non}{\nonumber}
\newcommand{\bquo}{\begin{quote}}
\newcommand{\enqu}{\end{quote}}
\def\de{\partial}
\def\Tr{ \hbox{\rm Tr}}
\def\H{ \hbox{\rm H}}
\def\Im{ \hbox{\rm Im}}
\def\Ker{ \hbox{\rm Ker}}
\def\const{\hbox {\rm const.}}
\def\o{\over}
\def\im{\hbox{\rm Im}}
\def\re{\hbox{\rm Re}}
\def\bra{\langle}\def\ket{\rangle}
\def\Arg{\hbox {\rm Arg}}
\def\Re{\hbox {\rm Re}}
\def\Im{\hbox {\rm Im}}
\def\exo{\hbox {\rm exp}}
\def\diag{\hbox{\rm diag}}
\def\longvert{{\rule[-2mm]{0.1mm}{7mm}}\,}
\def\a{\alpha}
\def\dag{{}^{\dagger}}
\def\tq{{\widetilde q}}
\def\p{{}^{\prime}}
\def\W{W}
\def\N{{\cal N}}
\def\hsp{,\hspace{.7cm}}
\newcommand{\Z}{\ensuremath{\mathbb Z}}
\newcommand{\vac}{\ensuremath{|0\rangle}}
\newcommand{\vact}{\ensuremath{|00\rangle}                    }
\newcommand{\oc}{\ensuremath{\overline{c}}}
\begin{titlepage}
\begin{flushright}
ULB-TH/05-26\\
hep-th/0512160\\
\end{flushright}
\bigskip
\def\thefootnote{\fnsymbol{footnote}}

\begin{center}
{\large {\bf
Killing Horizons
 } }
\end{center}

\bigskip
\begin{center}
{\large  Jarah EVSLIN\footnote{\texttt{ jevslin@ulb.ac.be}} 
 \vskip 0.10cm
 }
\end{center}

\renewcommand{\thefootnote}{\arabic{footnote}}

\begin{center}
International Solvay Institutes,\\
Physique Th\'eorique et Math\'ematique,\\
Universit\'e Libre
de Bruxelles,\\C.P. 231, B-1050, Bruxelles, Belgium\\
\vskip 2.3cm

\end {center}

\noindent
\begin{center} {\bf Abstract} \end{center}
QED in two-dimensional Minkowski space contains a single physical state as seen either by an inertial observer or by a constantly accelerating Rindler observer.  However in Feynman gauge if one takes a generic representative of the physical Minkowski state and traces over all left Rindler states, one does not arrive at the physical right Rindler state, but rather at a ``density matrix'' with negative eigenvalues for negative norm states corresponding intuitively to the radiation of uncorrelated temporal photons and ghosts.  This reflects the fact that states that are exact under the Minkowski BRST operator are not necessarily exact or even closed under the Rindler BRST operator.  Such situations are avoided when there are quantum corrections to the Hamiltonian that eliminate the horizons, which yield Mathurian fuzzball solutions.
\vfill

\begin{flushright}
\today
\end{flushright}
\end{titlepage}

\bigskip

\hfill{}
\bigskip

\section{Introduction}

\subsection{Philosophical Rambling on the Motivation for this Note}

This subsection is devoid of content, the reader may begin with Subsection~\ref{intro}.

Einstein's theory of gravity is nonrenormalizable \cite{Sagnotti1,Sagnotti2}.  This means that the cancellation of all divergences requires the introduction of an infinite number of independent counterterms, whose coefficients must then be measured.  An infinite set of measurements is not possible, and so the only hope of determining the coefficients in finite time is to hypothesize a property that they all satisfy and then test this property.  For example imposing Lorentz invariance upon the standard model has reduced the number of measurable parameters down to about 30, which aside from some neutrino mass matrix elements and the Higgs mass have been measured quite well.  Of course the standard model suffers from a Landau pole and a Higgs mass divergence, whose cures tend to lead people to the MSSM which has over 100 parameters (far more than one can hope to measure in the next generation of experiments), again leading to the need to guess a principle.

There are two promising candidates for principles in the literature.  One is that our universe is the target space of a two-dimensional conformal field theory (CFT).  If the CFT is invariant under target space diffeomorphisms then one appears to generally arrive at spacetime equations of motion that consist of Einstein's equation plus an infinite series of corrections which are the counterterms that cancel the UV divergences.  This principle is not sufficient to completely determine the coefficients, instead there is a set of counterterms for each CFT up to dualities.  It is also not minimal in that in addition to supplying the necessary infinite set of couplings it also leads to an infinite tower of massive excitations.  However it supplies some examples of quantum gravity theories with such high degrees of symmetries that otherwise unapproachable problems may be attacked, for example one can try to determine whether a black hole's horizon survives quantum corrections \cite{WittenCoset,CGHS,Lunin1,Lunin2,Mathur1,Mathur2,Mathur3,Mathur4,Mathur5,Mathur6,Mathur7}.  Whether or not our universe is the target space of a CFT, this class of theories is useful in that one may find general features of quantum gravity theories or at the very least test proposed quantum gravity no-go theorems.

In this note we will investigate a second candidate principle, proposed by 't Hooft \cite{thooft85,thooft94,thooft96,thooft05}, in which one attempts to find the restrictions on a quantum gravity theory imposed by unitarity in a background with horizons.  This principle will be used in the following form:

\noindent
{\textbf{Principle}} \textit{The universe as seen by any observer obeys the usual laws of quantum mechanics.  In particular the Hilbert space is positive definite.}

It is possible that this proposed principle places no constraint on the quantum completion of Einstein gravity.  For example it may be that all quantum completions of the relevant quantum gravity theories are holographic duals of unitary field theories and so must themselves be unitary.  However there are many different observers, at least as many as there are trajectories, and so it is also possible that this principle places an infinite number of constraints on the coefficients of the counterterms, maybe even enough to determine them.  At any rate, a stronger version of this principle is available which imposes that the various quantum theories are actually equivalent \cite{EP1,EP2,EP3}.  In the present note we will not be so ambitious.  Instead we will try to understand one of the simplest possible situations in which this principle may be nontrivial.

\subsection{The Problem and Five Possible Cures} \label{intro}
We will consider QED in two-dimensional Minkowski space, a topological theory with a single physical state for any given observer.  The extension to QED in a larger number of dimensions is trivial, one need only add the corresponding transverse oscillator modes which will commute with all of the two-dimensional oscillators.  The resulting Hilbert space factorizes into the 2d space discussed below, which contains a single physical state, tensored with the transverse excitations which are all physical.  Thus one need only replace the vacuum state in the following formulas with an arbitrary transverse excitation, and the arguments below will follow for each transverse excitation independently.  

We will argue that a representative of the physical state observed by an inertial observer does not always correspond to the physical state of a uniformly accelerating Rindler observer.  This is manifested in the fact that, using Feynman gauge in which the unphysical temporal polarization of the photon and the ghosts are free scalars, the most naive extension of the Unruh effect \cite{Unruh} predicts that the accelerating observer in the physical Minkowski vacuum will see a thermal spectrum of every excitation.  In particular the density matrix, which is not really a density matrix as it contains negative eigenvalues, will include negative norm states with odd numbers of photons with time-like polarization.  These will not be canceled in generic entries of the density matrix by ghosts, as the distributions of ghosts and temporal photons are thermal and uncorrelated.  

We will trace this apparent pathology to the fact that many representatives of the physical Minkowski state, which are all killed by the BRST charge integrated over a Cauchy surface, are not killed by the BRST charge integrated over a spatial slice of the Rindler wedge.  There is however one representative, which as always differs from the others by a BRST exact state, that is also killed on each wedge by the BRST charge and so its trace is the physical Rindler state.  This dependence on the chosen representative is possible because the Rindler and Minkowski BRST operators are different, the Minkowski BRST operator is the sum of the two Rindler operators.  Thus two Minkowski states which represent the same Minkowski BRST cohomology class do not necessarily trace to two Rindler states in the same Rindler BRST cohomology class.

There are at least five ways out of this situation.  First, one can drop the principle.  After all, we have no convincing argument that there should be a consistent quantum theory for the Rindler observer, nor that his states should be obtained by a trace.  One might also dismiss this problem as it rests upon the choice of the Feynman gauge, however if there is no gauge anomaly the theory should be consistent in any gauge.  On the other hand, the incompatibility of the physical state conditions may indicate an inability to enforce the gauge invariances of the Minkowski and Rindler observers simultaneously.

Second one may decide that the Rindler observer is unphysical.  After all the observer needs to accelerate forever in order to stay out of contact with the left Rindler wedge.  If at some point he finishes accelerating then he will reach timelike infinity and will be in the forward lightcone of every event.  The observer can only accelerate forever if a force acts upon him forever, which will require an infinite amount of energy from the viewpoint of an inertial observer.  If one includes the backreaction of this infinite energy on the geometry then one may conclude that the Minkowski space approximation is invalid, and so in the presence of backreaction this story would be radically changed.

Third, one could start with the Minkowski vacuum, which could be rewritten as an excitation of the tensor product of the right and left Rindler wedge vacua and then only trace over states in the left wedge for which the BRST charge in the left wedge vanishes.  This would leave the Rindler vacuum on the right wedge.  However we have no physical justification for such a selective trace.  After all many representatives of the Minkowski vacuum do not satisfy the Rindler physicality conditions and yet represent a legitimate physical state.  One may suspect that in an interacting theory such a projection onto the Rindler-physical spectrum would be illegal as the removed states would not decouple from an observable such as the S-matrix.  For example such a projection may lead one to miss the $-26R$ contribution to the stress tensor of Hawking radiation emitted by a 2-dimensional black hole \cite{ChristiansenFulling77}.  Of course there is no guarantee that the negative-normed states will decouple once the horizon is eliminated, although there will be no thermal Hawking radiation.

Fourth, one can mandate that the trace is only to be performed upon a preferred representative of the Minkowski physical state whose BRST charge vanishes separately on the interior and exterior of the horizon.  In other words, one may conclude that only certain gauges are suitable for performing the trace and Minkowski gauge transformations which take one out of this set of gauges are anomalous from the viewpoint of the Rindler observer.  We cannot show that such a representative exists in more general systems, but in the example at hand we will show that such a representative does exist and so this modification of the prescription is allowed.  Again in the case of the 2d black hole without 26 matter fields the negative Hawking radiation may imply that there is no such representative, as no gauge transform in the total spacetime can change the stress tensor.  Thus the existence of the representative may, as desired, place a restriction on the theories considered.  Notice however that the choice of a preferred representative appears to be in conflict with gauge invariance as the consistency of the Rindler observer's theory restricts the possible Minkowski gauge transformations.

At least in the example at hand, the preferred representative prescription is operationally equivalent to the selective trace prescription.  After all the rationale for only tracing over states that are physical behind the horizon is that the state behind the horizon should in some sense really be physical.  In effect one has imposed that the wave function behind the horizon is annihilated by the BRST operator behind the horizon, but only the preferred representative wave function has this property.  The ordinary trace of the preferred representative equals the selective trace of an arbitrary representative.  

The final way out is to say that geometries with horizons, while they are solutions to the classical equations of motion, are not solutions to the full quantum-corrected equations of motion.  This would be an extension of the Callan, Giddings, Harvey and Strominger proposal that black holes evaporate before they can form \cite{CGHS}.  In string theory this would be a generalization of Mathur's conjecture \cite{Mathur1}.  In two-dimensional black holes there also appears to be some support for such a conjecture.  In Ref.~\cite{WittenCoset} Witten has found that for a black hole described by an exactly solvable CFT, albeit in an approximation, the minimum radius that an infalling shell of spherical matter may reach is greater than the Schwarzschild radius.  The motivation for the present work came from Ref.~\cite{Strom} in which Strominger presents evidence that the stress tensor of matter is precisely canceled by a quantum correction as the matter approaches the event horizon.  There the author removes the stress-energy contribution of the ghosts by hand by adding a counterterm.  No such counterterm will be included in the present note.

Following a suggestion of D.~Anselmi and M.~Henneaux, in Section~\ref{brstsec} we will review the construction of the BRST operator in Minkowski space, which is the integral of a local function.  It may be written as the sum of the BRST operators on the two Rindler wedges, which correspond to the integral of the local function over the two Rindler halves of a Cauchy surface.  Next in Section~\ref{fantasmi} we will show that a representative of the Minkowski vacuum is annihilated by the Minkowski BRST charge but not by either of its Rindler summands, which instead annihilate the Rindler vacuum. We will write the Minkowski vacuum as a sum of tensor products of left and right Rindler states and then trace out the left ones to reproduce what the accelerating observer might see and we will find that he sees dead people, as well as nongauge-invariant, negative-normed, time-like polarized photons.  Of course, as noted above, a trace over only left-physical states would eliminate the pathology.  Finally in Sec.~\ref{sogni} we note that such problems also go away if one eliminates horizons, as does, for example, the horizon problem.  While inflation provides a temporary solution to the horizon problem\footnote{That is, if we use inflation to solve the horizon problem then someday we will see CMB from matter that was never in causal contact and it may have a strange temperature profile.} the lack of horizons would provide a permanent, albeit far-fetched solution.  We note that quantum corrections that eliminate horizons also tend to preserve neither Lorentz-invariance nor general covariance, but who does?

\section{The BRST Operators} \label{brstsec}

\subsection{A Review of the Minkowski BRST Charge} \label{mink}

We will be interested in 2-dimensional QED in Minkowski space, which is described by the action
\beq
S=\frac{1}{4}\int dxdt\ ((\partial_0A_1)^2-2\partial_0A_1\partial_1 A_0+(\partial_1A_0)^2) \label{azione}
\eeq
where $A_\mu$ is the photon field.  The action is independent of the time derivative of $A_0$ and so $A_0$ is not a dynamical field, as its equation of motion allows it to be obtained on a given time slice from only the boundary conditions on the timeslice and derivatives of $A_1$.  The action (\ref{azione}) enjoys a gauge invariance $A_\mu\mapsto A_\mu+\partial_\mu\lambda$ which can be used to cancel the $A_1$ degree of freedom, for example by choosing the axial gauge $A_1=0$, and so the gauge-fixed equations of motion yield no dynamical degrees of freedom.  In the quantum theory this is reflected, as we will review in Sec.~\ref{fantasmi}, in the existence of a unique physical state.

The action (\ref{azione}) is not suitable for a path-integral formulation of the theory because the kinetic term is not invertible and so the propagator is not defined.  To produce a suitable action we will add a gauge-dependent term, which will introduce an undesirable determinant factor in the path integral which we will cancel via the addition of ghosts.  Our choice of gauge will not affect the physics so long as we restrict our attention to gauge-invariant states, which will be the case if we consider only BRST-closed states.  However we will argue that some Minkowski gauge choices lead to nonphysical Rindler states.  Thus our choice of gauge may affect the physics of those observers, for example by immersing them in a thermal bath of unphysical particles.  Of course we do not want to argue that one can really observe either the gauge choice or the unphysical particles, but rather the goal is to argue that there is a problem with our naive adaptation of a standard formalism to gauge fields.

We will choose the Feynman gauge, in which $A_0$ and $A_1$ are free scalars whose contributions to the path integral are precisely canceled by the ghost $c$ and antighost $\oc$, which are also free scalars,
\beq
S=\int dxdt\ (-\frac{1}{4}\partial_\mu A_0\partial^\mu A_0+\frac{1}{4}\partial_\mu A_1\partial^\mu A_1-\partial^\mu \oc\partial_\mu c) \label{feyn}
\eeq
because the ghost and antighost have fermionic statistics.  As the contributions to the path integral measure of all of the fields cancel, the measure is just one, leading to an effective action which is zero.  Thus it seems implausible that any observer will observe any kind of excitation in this theory, since the path integral is trivial and would be unchanged if all of the fields were removed.  In fact a simpler example that could have been used in this note instead of QED is a pair of free bosons with opposite statistics, which also yields a trivial path integral.  

The classical equations of motion for the various fields may be obtained by varying the action (\ref{feyn}).  One finds the Klein-Gordon equation for all four scalars, whose solutions may be Fourier decomposed into plane waves.  The BRST operator will be constructed from the ghost field $c$ and the longitudinal photon $k^\mu A_\mu$ where $k^{\mu}$ is the 2-momentum.  In momentum space the Klein-Gordon equation becomes the mass shell condition $|k_0|=|k_1|$.  If we define the frequency $w=|k_0|$ then we may normalize the longitudinal photon as follows
\beq
A=A_0+\frac{k}{\omega}A_1\hsp k=k^1\hsp w=|k^0|\geq 0.
\eeq
We may now decompose the ghost and longitudinal photon
\bea
A(x,t)&=&\int\frac{dk}{\sqrt{4\pi\omega}}(a_ke^{ikx-i\omega t}+a^\dagger_ke^{-ikx+i\omega t})\nonumber\\
c(x,t)&=&\int\frac{dk}{\sqrt{4\pi\omega}}(c_ke^{ikx-i\omega t}+c^\dagger_ke^{-ikx+i\omega t}).
\eea

The coefficients $a_k$, $a^\dagger_k$, $c_k$ and $c^\dagger_k$ can be used to construct the Minkowski space BRST operator
\beq
Q_M=\int dk[c^\dagger_ka_k+c_ka^\dagger_k] \label{mbrst}
\eeq
which may be written as an integral over a Cauchy surface of a local charge density $\rho$
\beq
Q_M=\int_{-\infty}^{\infty} dx \rho(x,t)\hsp \rho(x,t)=iA(x,t)\frac{\partial}{\partial t} c(x,t)-ic(x,t)\frac{\partial}{\partial t} A(x,t). \label{rho}
\eeq
Ghostnumber zero states which are annihilated by $Q_M$ will automatically be gauge invariant, or more precisely will be invariant under positive-frequency gauge transformations as one sees explicitly in the Gupta-Bleuler quantization of electrodynamics.  Had we used Gupta-Bleuler instead of BRST we would have found that the Minkowski and Rindler states are invariant under different sets of gauge transformations.

Unruh has argued \cite{Unruh} that positive-frequency corresponds to holomorphic in our flat two-dimensional example.  This led him to the conclusion that one may substitute the plane wave basis with any holomorphic basis of Klein-Gordon solutions in the definition of the vacuum state.  While he was discussing theories of free scalars, the same argument applies here.  One may replace the BRST operator in Eq.~(\ref{mbrst}) with another in which the plane wave decomposition has been replaced with another holomorphic decomposition and the coefficients of the two terms are arbitrary and one will find the same BRST cohomology representing the same physical states.  We will find that it will be computationally easier to work with a different decomposition in the following subsection.

\subsection{A Quick Review of Rindler Space}
In general relativity there are always observers, even in Minkowski space, who do not ever have access to the entire spacetime because they accelerate sufficiently to never enter the forward (backward) lightcones of some events.  The boundary of the events whose forward (backward) lightcones are intersected by a particular trajectory is a future (past) horizon of the observer following that trajectory.  In particular an observer in 2d Minkowski space who experiences a constant proper acceleration for all time will have a past and a future horizon, which partition the spacetime into four quadrants $F, P, L$ and $R$.  We will choose the origin such that these regions are defined by $t>|x|$, $t<-|x|$, $x<-|t|$ and $x>|t|$ respectively.  While no single region contains a Cauchy surface, the union of the left and right regions $L$ and $R$ plus the origin does contain many Cauchy surfaces, for example $t=0$.  States correspond to wave functions on Cauchy surfaces, and so it will be possible to describe some states once we describe $L$ and $R$.

We will write the Minkowski metric with the sign convention
\beq
ds^2=dt^2-dx^2.
\eeq
In the quadrant $R$, called the Rindler wedge or the right Rindler wedge, one may define the coordinates $(\xi^R,\eta^R)$ via the transformations
\beq
t=\frac{e^{a\xi^R}}{a}\textup{sinh}\ a\eta^R\hsp x=\frac{e^{a\xi^R}}{a}\textup{cosh}\ a\eta^R\hsp \label{rcoord}
\eeq
where $a$ is a dimensionful constant.  An observer with constant proper acceleration $g>0$ remains at a constant $\xi^R$ coordinate 
\beq
\xi^R=\frac{ln(a/g)}{a}.
\eeq
We may define coordinates $(\xi^L,\eta^L)$ in the left Rindler wedge $L$ using Eq.~(\ref{rcoord}) but with the signs of both $t$ and $x$ reversed.  

In these new coordinates the metric is conformal to the Minkowski metric
\beq
ds^2=e^{2a\xi}(d\eta^2-d\xi^2)
\eeq
and so the positive frequency plane waves will still be of the form $e^{ik\xi^R-i\omega\eta^R}$\ and $e^{ik\xi^L+i\omega\eta^L}$.  The sign difference corresponds to the fact that $\eta^L$ is monotonically decreasing with respect to $t$.  More generally, the usual correspondence between right-moving and positive frequency is reversed in the $(\xi^L,\eta^L)$ basis.  No linear combination of these two plane waves is holomorphic at the origin, however the sum of the plane wave on one side and $e^{-\pi\omega/a}$ times the conjugate planewave with negated wavenumber on the other side is everywhere holomorphic.

There are then three useful momentum decompositions for a Klein-Gordon field $A$, one may decompose in the basis of Subsection~\ref{mink}, in terms of the $(\xi,\eta)$ plane waves with support in $R$ and $L$, or in terms of the everywhere holomorphic linear combinations
\bea
A&=&\int\frac{dk}{\sqrt{4\pi\omega}}(a_ke^{ikx-i\omega t}+a^\dagger_ke^{-ikx+i\omega t})\label{decomps}\\
&=&\int\frac{dk}{\sqrt{4\pi\omega}}(a^L_ke^{ik\xi^L+i\omega \eta^L}+a^{L\dagger}_ke^{-ik\xi^L-i\omega \eta^L}+a^R_ke^{ik\xi^R-i\omega \eta^R}+a^{R\dagger}_ke^{-ik\xi^R+i\omega \eta^R})\nonumber\\
&=&\int\frac{dk}{\sqrt{4\pi\omega(e^{\pi\omega/a}-e^{-\pi\omega/a})}}[a^1_k(e^{\frac{\pi\omega}{2a}+ik\xi^R-i\omega\eta^R}+e^{\frac{-\pi\omega}{2a}+ik\xi^L-i\omega\eta^L})\nonumber\\
&&+a^2_k(e^{\frac{\pi\omega}{2a}+ik\xi^L+i\omega\eta^L}+e^{\frac{-\pi\omega}{2a}+ik\xi^R+i\omega\eta^R})
+a^{1\dagger}_k(e^{\frac{\pi\omega}{2a}-ik\xi^R+i\omega\eta^R}+e^{\frac{-\pi\omega}{2a}-ik\xi^L+i\omega\eta^L})\nonumber\\
&&+a^{2\dagger}_k(e^{\frac{\pi\omega}{2a}-ik\xi^L-i\omega\eta^L}+e^{\frac{-\pi\omega}{2a}-ik\xi^R-i\omega\eta^R})]\nonumber
\eea
where the plane waves $e^{ik\xi^L+i\omega \eta^L}$ are defined to be zero in the wedge $R$ and similarly the other waves vanish in the wedge $L$.  The ghost field $c$ will also be decomposed as in (\ref{decomps}).  Matching coefficients in the second and third expression one finds the Bogoliubov transforms
\beq
a^L_k=\frac{e^{-\pi\omega/2a}a^{1\dagger}_{-k}+e^{\pi\omega/2a}a^2_k}{\sqrt{e^{\pi\omega/a}-e^{-\pi\omega/a}}}\hsp
a^R_k=\frac{e^{-\pi\omega/2a}a^{2\dagger}_{-k}+e^{\pi\omega/2a}a^1_k}{\sqrt{e^{\pi\omega/a}-e^{-\pi\omega/a}}}. \label{bog}
\eeq

\subsection{The BRST Charge in a Rindler Wedge}
In principle one could calculate the BRST charge operators $Q_L$ and $Q_R$ on the Rindler wedges $L$ and $R$ by integrating the quantity $\rho$ from Eq.~(\ref{rho}) over $x<0$ and $x>0$ respectively.  While this is not difficult, an alternate strategy is to guess a local BRST charge density that gives a simple answer for $Q_L$ and $Q_R$ in terms of the Rindler mode expansions above and then show that the sum $Q_L+Q_R$ is a legitimate Minkowski space BRST operator, that is that $Q_L+Q_R$ has the same cohomology as $Q_M$ or equivalently that it has the same form as in (\ref{mbrst}) in terms of holomorphic variables with arbitrary coefficients as described above.  We will follow this alternate strategy so that the Rindler operators will be expressed as simply as possible.

We will guess that the left and right wedge BRST charges are
\beq
Q_L=-\int dk[c^{L\dagger}_ka^L_k+c^L_ka^{L\dagger}_k]\hsp
Q_R=\int dk[c^{R\dagger}_ka^R_k+c^R_ka^{R\dagger}_k]. \label{qguess}
\eeq
Note that these charges produce the usual physical spectrum in the two Rindler wedges.  In particular there will be a single state in each wedge, the vacuum, which will be annihilated by all of the lowering operators in the $(L,R)$ basis.  The relative sign difference between the operators does not affect their cohomologies, but will affect the cohomology of their sum.  Notice that, like $Q_M$, the charges $Q_L$ and $Q_R$ can both be expressed as the integrals of local quantities.  In fact the local quantity is defined identically to $\rho$ in (\ref{rho}) but using the basis $(\xi,\eta)$.

Adding $Q_L$ and $Q_R$ we obtain a candidate Minkowski space BRST operator
\bea\label{ql}
Q_L+Q_R&=&\int dk[-c^{L\dagger}_ka^L_k-c^L_ka^{L\dagger}_k+c^{R\dagger}_ka^R_k+c^R_ka^{R\dagger}_k]\\
&=&\int\frac{dk}{e^{\pi\omega/a}-e^{-\pi\omega/a}}[-(e^{-\pi\omega/2a}c^{1}_{-k}+e^{\pi\omega/2a}c^{2\dagger}_k)(e^{-\pi\omega/2a}a^{1\dagger}_{-k}+e^{\pi\omega/2a}a^2_k)\nonumber\\
&&-(e^{-\pi\omega/2a}c^{1\dagger}_{-k}+e^{\pi\omega/2a}c^{2}_k)(e^{-\pi\omega/2a}a^{1}_{-k}+e^{\pi\omega/2a}a^{2\dagger}_k)\nonumber\\
&&+(e^{-\pi\omega/2a}c^{2}_{-k}+e^{\pi\omega/2a}c^{1\dagger}_k)(e^{-\pi\omega/2a}a^{2\dagger}_{-k}+e^{\pi\omega/2a}a^1_k)\nonumber\\
&&+(e^{-\pi\omega/2a}c^{2\dagger}_{-k}+e^{\pi\omega/2a}c^{1}_k)(e^{-\pi\omega/2a}a^{2}_{-k}+e^{\pi\omega/2a}a^{1\dagger}_k)]\nonumber\\
&&=\int dk[+c^{1\dagger}_ka^1_k+c^1_ka^{1\dagger}_k-c^{2\dagger}_ka^2_k-c^2_ka^{2\dagger}_k].\nonumber
\eea
As the $(1,2)$ basis is holomorphic, the final equality in Eq.~(\ref{ql}) implies that $Q_L+Q_R$ annihilates the same states as $Q_M$ and in fact provides the same physical state condition.  Thus we may start with $Q_L+Q_R$ as the definition of our Minkowski BRST operator, which leads to $\rho$ defined using $\partial_{\pm\eta}$ instead of $\partial_t$, whose integrals over negative and positive $x$ give $Q_L$ and $Q_R$ respectively.

Notice that neither set of Rindler coordinates covers any of the horizons, and that the Rindler plane waves are not continuous at the horizons using the Minkowski topology as the wavenumbers go to infinity in Minkowski coordinates.  This means that Minkowski-continuous Rindler planewaves must vanish at the horizons, and so perhaps neither $Q_L$ nor $Q_R$ may be used to impose gauge invariance at the horizon.  The horizon is an important point on each Cauchy surface, as all of the unphysical particle radiation entering the right Rindler wedge passes through the horizon.  It is not so surprising that an observer who is unable to impose gauge invariance on a hypersurface is bombarded with nongauge-invariant radiation from that hypersurface.

\section{The States: It's Always Halloween on the Rindler Wedge} \label{fantasmi}

\subsection{The Spectrum of QED in Minkowski Space}
The Minkowski observer may construct his Hilbert space as follows.  The available momentum space operators are the Fourier modes of the photon polarizations $A_0$ and $A_1$ as well as the ghost $c$ and the antighost $\bar{c}$.  Without loss of generality we will take $k>0$ and define a longitudinal photon $A$ and temporal photon $B$ by
\beq
A=A_0+A_1\hsp B=A_0.
\eeq
The modes will be denoted $a_k,\ b_k,\ c_k,\ \oc_k,\ a_k^\dagger,\ b_k^\dagger,\ c_k^\dagger$,\ and $\oc_k^\dagger$ and satisfy the usual harmonic oscillator relations
\beq
-[a_k,b^\dagger_{k\p}]=-[b_k,b^\dagger_{k\p}]=-[c_k,\oc^\dagger_{k\p}]_+=[\oc_k,c^\dagger_{k\p}]_+=\delta(k-k\p) \label{alg}
\eeq
where all other commutators are zero.  Even $[a_k,a_{k\p}^{\dagger}]$ vanishes because it is the sum of the commutators of the timelike and spacelike generators which differ by a sign.  He begins with the vacuum state $\vac_M$, which is defined by the relations
\beq
a_k\vac_M=b_k\vac_M=c_k\vac_M=\oc_k\vac_M=0. \label{vacdef}
\eeq
The Hilbert space consists of polynomials in the creation operators $a_k^\dagger,\ b_k^\dagger,\ c_k^\dagger$,\ and $\oc_k^\dagger$ acting on the vacuum.  It admits a nonnegative grading given by the degree of the polynomial.

The physical states are the ghost number zero cohomology of the operator
\beq
Q_M=\int dk[c^\dagger_ka_k+a^\dagger_kc_k]
\eeq
from Eq.~(\ref{mbrst}).  In particular $Q_M$ preserves the degree of a state.  At degree zero the only state in the Hilbert space is $\vac_M$.  It is annihilated by both $a_k$ and $c_k$ and so it is $Q_M$-closed.  As $Q_M$ preserves the grading, and it annihilates the only degree zero state, there can be no degree zero states in its image.  Thus $\vac_M$ is not $Q_M$-exact and, being ghost number zero, it is therefore a physical state.

An arbitrary state at degree one is
\beq
|\psi\rangle=\int dk(\alpha(k)a^\dagger_k+\beta(k)b^\dagger_k+\gamma(k)c^\dagger_k+\overline{\gamma}(k)\oc^\dagger_k)\vac_M.
\eeq
Acting with the BRST operator $Q_M$ on $|\psi\rangle$ one finds
\beq
Q_M|\psi\rangle=-i\hbar\int dk(\beta(k)c^\dagger_k+\overline{\gamma}(k)a^\dagger_k)\vac_M.
\eeq
The Hilbert space generators $c^\dagger_k\vac_M$ and $a^\dagger_k\vac_M$ are seen to be in the image of $Q_M$, while no combination of $b^\dagger_k\vac_M$ and $\oc^\dagger_k\vac_M$ is in its kernel.  Thus the BRST cohomology at degree one is
\beq
\frac{\Ker(Q_M)}{\Im(Q_M)}=\frac{\int dk(\alpha(k)a^\dagger_k+{\gamma}(k)c^\dagger_k)\vac_M}{\int dk(\alpha(k)a^\dagger_k+{\gamma}(k)c^\dagger_k)\vac_M}=0
\eeq
where $0$ denotes the trivial group, which consists of only one element.  In other words there are no physical states at degree one.  Generalizing the above argument to arbitrary polynomials is straightforward and one finds that there are no physical states at any higher degree.

Instead of using a Fourier decomposition of states one could have used the holomorphic $(1,2)$ basis.  As the generators satisfy two commuting copies of the same algebra (\ref{alg}), one would find the same BRST cohomology.  The fact that the destruction operator in the Fourier basis may be decomposed into only destruction operators in the $(1,2)$ basis implies that the lone physical state, the vacuum,  is the same in both quantization schemes.  This is because the definition of the vacuum (\ref{vacdef}) may be rewritten, using this decomposition, as the definition of the $(1,2)$ basis vacuum.

\subsection{The Minkowski Vacuum as Seen by A Rindler Observer}

The above argument may also be used on the right (left) Rindler wedge using the operator $Q_R$ ($Q_L$) instead of $Q_M$ and one would find a unique state, the Rindler vacuum $\vac_R$ ($\vac_L$) defined by
\beq
a^R_k\vac_R=b^R_k\vac_R=c^R_k\vac_R=\oc^R_k\vac_R=0. 
\eeq
This state is a wavefunction on a spatial slice of the Rindler wedge, not on a Cauchy surface like the descendants of $\vac_M$, and so the vacua do not inhabit the same Hilbert space.

However one may construct a Cauchy surface in Minkowski space by attaching a spatial surface on each Rindler wedge that ends at $\eta=0$ to the origin.  If we ignore the fact that neither Rindler surface contains the origin, this implies that the Minkowski Hilbert space is the tensor product of the left and right Rindler Hilbert spaces.  As the operators in the $(L,R)$ basis correspond to decompositions with support in only one wedge, the tensor product with the other wedge will not change their action.  For example
\beq
a^R_k\vact=b^R_k\vact=a^L_k\vact=b^L_k\vact=0\hsp\vact=\vac_L\otimes\vac_R 
\eeq
and similarly the ghost annihilation operators destroy the tensor product vacuum.

This does not imply that the $(1,2)$ basis annihilation operators destroy the tensor product vacuum.  In fact they do not, as in Eq.~(\ref{bog}) we saw that they contain $(L,R)$ destruction operators which kill the vacuum, plus $(L,R)$ creation operators that are linearly independent and do not kill it.  Thus the tensor vacuum is not the same element of the Hilbert space as the Minkowski vacuum, instead it is a squeezed state
\beq
\vac_M=\exp[\int dk e^{-\frac{\pi\omega}{a}}(d^{L\dagger}_ka^{R\dagger}_k+a^{L\dagger}_kd^{R\dagger}_k+c^{L\dagger}_k\oc^{R\dagger}_k-\oc^{L\dagger}_kc^{R\dagger}_k)]\vact \label{urel}
\eeq
where we have fixed the arbitrary relative normalization and $d_k=a_k/2-b_k$ are the modes of the unphysical photon $(A_1-A_0)/2$.  Notice that both sides of (\ref{urel}) are annihilated by the $(1,2)$ basis destruction operators.  The exponent is ghost number zero and so both $\vac_M$ and $\vact$ are ghost number zero.  $\vact$ is annihilated by $Q_L$ and $Q_R$ separately and so it is annihilated by the sum, which is a choice of Minkowski BRST operator.  $\vac_M$ we have argued is also annihilated by any Minkowski BRST operator.  As, up to normalization, there is only one element in the Minkowski BRST cohomology we conclude that $\vact$ and $\vac_M$ are two representatives of the same BRST cohomology class.  In fact a short calculation shows that they indeed differ by a BRST exact state
\beq
\vac_M-\vact=(Q_L+Q_R)\int dk e^{-\frac{\pi\omega}{a}}(\oc^{L\dagger}_kd^{R\dagger}_k-d^{L\dagger}_k\oc^{R\dagger}_k)\frac{e^\eta-1}{\eta}\vact \label{diff}
\eeq
where $\eta$ is the exponent in Eq.~(\ref{urel}).  Note that while $\eta$ is not invertible, the $\eta$ in the denominator of (\ref{diff}) is canceled a factor of $\eta$ in the numerator.  Thus we may represent the Minkowski vacuum by either $\vact$ or $\vac_M$, the difference corresponding roughly to a Minkowski gauge choice.  As we are looking for trouble, we will consider $\vac_M$.

To avoid writing infinite sums, we will pick out two particular terms on the right hand side that we will follow
\beq
\vact-e^{\pi\omega/a}b^{L\dagger}_kb^{R\dagger}_k\vact. \label{pezzo}
\eeq
It will be important that $b^{L\dagger}_kb^{R\dagger}_k\vact$ is the only term in the sum that consists of a left state tensored with $b^{R\dagger}\vac_R$.

Now we come to the crucial question.  We know that QED is consistent in Minkowski space, and so $\vac_M$ is a very good physical state.  We will consider the universe in this state.  We want to know what state an eternally rightwards accelerating observer will see.  Of course no such observer can really exist, so it is an abstract question and there is no convincing argument that it should have an answer.  However we know that in general relativity this observer has a horizon, and if he is trying to define a state on the surface $t=\eta=0$ then he will only have access to the state in the right Rindler wedge $x>0$.
That is, if we decompose the state as a sum of tensor products of left and right Rindler states, then he will only be able to measure the right parts.  In particular he will not know how the left parts are correlated, and so instead of seeing a pure state, one may conclude that he only has access to enough information to construct a density matrix.  The density matrix obtained from a Minkowski state by forgetting about the left components is the trace of the Minkowski state over the left Hilbert space.

This motivates the most naive possible definition of the state observed by our accelerating observer as the density matrix obtained by a left trace of $\vac_M$.  The trace is a linear operation, and so we may perform it term by term.  For example the trace of the $\vact$\ term gives the right vacuum $\vac_R$.  As $\vact$\ is the only term on the right side of (\ref{urel}) that is the product of a left state with $\vac_R$, this will be the only contribution to the density matrix term $\vac_R{}_R\langle 0|$.  We may add the trace of the $b^{L\dagger}_kb^{R\dagger}_k\vact$ term from Eq.~(\ref{pezzo}), to obtain
\beq
\Tr_L(\vact-e^{-\frac{\pi\omega}{a}}b^{L\dagger}_kb^{R\dagger}_k\vact)(\langle 00|-e^{-\frac{\pi\omega}{a}}\langle 00|b^{L}_kb^{R}_k)
=\vac_R{}_R\langle 0|-e^{-\frac{2\pi\omega}{a}}b^{R\dagger}_k\vac_R{}_R\langle 0|b^{R}_k \label{tpic}
\eeq
where the minus sign on the right hand side comes from the wrong sign temporal photon commutation relation
\beq
{}_L\langle 0|b^{L}_kb^{L\dagger}_{k\p}\vac_L=i\hbar\delta(k-k\p).
\eeq
Notice that if we had instead chosen $\vact$ as our vacuum representative we would have found that the trace is just $\vac_R$.

The trace of interest, $\Tr_L\vac_M{}_M\langle0|$, will consist of the two terms in Eq.~(\ref{tpic}) plus other linearly independent terms.  The other terms are linearly independent because the decomposition (\ref{urel}) does not contain any other states that are the tensor of a left state times either the right vacuum or a single right temporal photon.  Thus there will be no other terms that will cancel the pathologies of the trace in Eq.~(\ref{tpic}).  In particular the trace will not be a density matrix, as it contains terms, such as $b^{R\dagger}_k\vac_R{}_R\langle 0|b^{R}_k$, with negative eigenvalues.  This ruins the probabilistic interpretation of the density matrix.  In addition the terms with negative eigenvalues are negative norm states, which conflicts with the principle that we have hypothesized in the introduction.  Only one element of the density matrix, $\vac_R{}_R\langle 0|$, is the physical Rinder state as measured by $Q_R$.  Thus the physical state condition is not compatible with the trace operation in the sense that the trace of a given representative of an $M$-physical state is not always $R$-physical.

Notice that the ghosts do not help matters as they are uncorrelated with the temporal photon.  In fact, in some elements of the density matrix, such as the second in (\ref{tpic}), they are not excited and so cannot cancel the temporal photon's deviant behavior element by element in the matrix.  The cancellation would need to be element by element as a single measurement places one in a pure state corresponding to any element in the diagonal density matrix, and there is a positive probability for measurements leading to unphysical states such as, for example, any even number of temporal photon excitations.

This incompatibility should not come as a surprise.  The state $\vac_M$ is physical because the integral of the BRST charge density $\rho$ over the entire $x$ axis kills it, that is, it carries no net BRST charge.  However $\rho$ does not kill it locally.  On the contrary, $Q_L$ and $Q_R$ are integrals of BRST charges over half of space and when expressed in the $(1,2)$ basis by inserting (\ref{bog}) into (\ref{qguess}) they contain terms of the form $a^{1\dagger}a^{2\dagger}$, which do not annihilate $\vac_M$.  Thus while an inertial observer in $\vac_M$ observes a universe with no net BRST charge, there is BRST charge density everywhere and an observer with access to only part of the universe will see a net charge.

\subsection{Is This Pathology Observable?}

We have argued that tracing the Minkowski vacuum in all gauges but one leads to a density matrix with negative-normed states, which violates our hypothesis that physical states should have a positive norm.  This leaves us with two possibilities, either the theory is unphysical, or else the hypothesis is too strong to be a constraint on the set of physical theories.  As an example of the second possibility, it may be that the Rindler observer cannot detect the unphysical modes in the density matrix, as one would expect since the two Minkowski vacua are gauge equivalent.

In support of this second possibility, expectation values computed using the physical density matrix, that obtained by tracing the tensor product vacuum, in fact agree with those calculated using an unphysical density matrix.  For example, the contributions of ghosts and longitudinal photons to the energy cancel each other.  The problem arises when the Rindler observer attempts to measure in which element of the density matrix he lies, for example, if he attempts to measure the number of temporal photons.  One may object that such an experiment may not be performed in a gauge-invariant fashion.  But this is precisely the point, by measuring the number of temporal photons the observer may gain information about the choice of gauge and so violate gauge invariance.

The expected number of temporal photons, which one finds by calculating the number of temporal photons in each density matrix element and then averaging by the respective eigenvalue, is zero.  Thus no information about the gauge is contained in this expectation value.  This does not imply however that the number of temporal photons measured by an observer will be zero.  On the contrary such a measurement collapses the density matrix of the mixed state to a smaller density matrix containing only those states for which the temporal photon number is equal to the measured value.  Likewise by measuring the number operator of all of the fields, the observer collapses the mixed state to a pure state.  

One may hope that in this pure state the contributions of the unphysical and gauge degrees of freedom will cancel.  However they do not in general cancel as the excitation numbers of the various fields are thermal and uncorrelated.  Thus the final pure state, which corresponds to a generic matrix element in the original mixed state, will contain different numbers of the various unphysical excitations.  For example, there is one possible final pure state which contains just a temporal photon and neither ghosts nor longitudinal photons.  Thus for this final pure state there is no hope that, for example, the energy of the temporal photon will be canceled by other unphysical degrees of freedom.

In conclusion, expectation values computed using gauge-equivalent density matrices are equal because the contributions of the unphysical degrees of freedom cancel each other as usual.  However, through measurements the observer can reduce the density matrix to a pure state, and the configurations of the various unphysical fields in the pure state are uncorrelated.  Thus in general these cancellations do not apply to the final pure state.  This leaves a number of possible resolutions.  One may choose to either accept the unphysical degrees degrees of freedom in the final pure state as physical.  One may reject the probabilistic interpretations of these density matrices, which after all have negative eigenvalues.  One may conclude that number operators of the various fields are not observable.  Or, finally, one may reject the classical analysis of Rindler space that was used above, as will be the subject of the following section.

\section{A World With No Horizons} \label{sogni}
What has this exercise taught us about the consistency of quantum gravity theories?  Perhaps it was a just a long derivation of the technical point that we need to choose our gauge correctly if we want to use the trace prescription to find the state observed by an observer with access to only part of the spacetime.  In the present example this corresponds to imposing two physical state conditions
\beq
Q|\psi\rangle=Q_L|\psi\rangle=0
\eeq
which are both satisfied only by the tensor product representative of the vacuum $\vact$.

However if we want to understand how various principles restrict the set of possible quantum gravity theories, then there are applications to two conjectured principles.  First, if one conjectures a positive definite Hilbert space for each observer, including the Rindler observer, then there is either a restriction on the set of allowed gauges or else the trace is not allowed.  If there is a restriction on the allowed gauges then it would be interesting to find the obstruction to the existence of a gauge in which the trace which respect to any subsector is physical.  For example the obstruction could be an element of the BRST cohomology at positive ghost number in which case such a gauge would always exist when the theory is anomaly free, but it could also be that the vanishing of this obstruction places a new restriction on quantum gravity theories.  If instead the trace is not allowed, we define a horizon in the quantum theory as a slice beyond which an observer needs to trace out the state, and so there would be no horizon in this quantum mechanical sense. 

Second, one may conjecture the principle that the Minkowski gauge transformations and in particular the addition of a Minkowski BRST exact state do not affect the positivity of the Hilbert space of any observer.  In this case one is left with the second solution, that there can be no horizon in the quantum theory.

Again we will degenerate into shameless speculation.

So what does it mean for there to be no horizons in the quantum theory?  We know that the Rindler observer in classical general relativity does have a horizon.  Thus it means intuitively that in quantum gravity theories somehow quantum processes allow him to probe the other side of the horizon, in contrast with the usual situation in which cross-horizon correlators fall off exponentially.  It is a common occurrence these days that pathologies in general relativity are cured in the quantum theory.  Not only can singularities be removed, but Gimon and Ho{\v r}ava have even shown that in string theory a generalized supertube may invalidate a classical solution near the boundary between a region with and without closed timelike curves and so effectively cut out the undesirable other side of the wall \cite{Horava}.  In a string theory context the replacement of the horizon and its interior by a membrane is an old idea.

The simplest mechanism for horizon elimination is a quantum correction to Einstein's equation which modifies the solutions such that the horizon goes away.  The physical effect of such a correction could be analogous to Mathur's solutions in the D1-D5 system, in which the singularity inside of a black hole is smeared so that the characteristic distance is of the same order as the Schwarzschild radius.  In these solutions there is not enough matter inside of any sphere for that sphere to be a Schwarzschild radius, and so there is no horizon.  While this derivation used the degrees of freedom of the worldvolumes of the D1 and D5-branes themselves (well, their U-duals), it may be that the same solution could arise from spacetime equations of motion with $\hbar$ corrections that come from, for example, the effective action.

What kind of corrections may eliminate horizons?  Horizons occur when $g_{00}/g_{kk}=0$ for any $k$.  Notice that this condition is not Lorentz-invariant, but there are several kinds of terms that may create such a restriction.  One may use the curvature of a spatial slice and its second fundamental form, and also one may use the determinant of the spatial part of the metric in the Hamiltonian, as the Hamiltonian already contains a hidden factor of $g_{00}$.  The latter term has the correct dimensions once one adds some factors of the Planck length, and it may be interpreted as a kind of Planck cube density.  It may, unfortunately, allow for Kasner-like solutions with horizons.  It does not appear, however, that the usual Lorentz-invariant polynomials in derivatives of $R$ that appear as counterterms to the Einstein-Hilbert action can eliminate horizons.  

A spatial Planck cube density is neither Lorentz-invariant nor generally covariant.  However it is not such a novel quantity.  Being a number, the entropy of a black hole is absolutely invariant in the quantum theory and yet in the absence of derivative corrections to the action it is a Planck area.  

The lack of Lorentz-invariance and general covariance in horizon-killing terms does not exclude them.  After all the universe is not Lorentz-invariant, the microwave background picks out a frame and even exerts a pressure on all matter (although more so on electrically charged matter) driving it to an Aristotelian rest with respect to its frame.  This is the frame of the last scattering surface, and so presumably the frame of the big bang.  Thus Lorentz-invariance in quantum gravity is at best a spontaneously broken symmetry, and so one expects the dynamical generation of Lorentz-dependent counterterms.

\subsection{Neutron Stars and Inflation}

We will stress that while the elimination of horizons is a possibility which may be the consequence of a hypothesized principle, we have presented no evidence that either this possibility or this principle is realized.\footnote{Indeed the goal of the program is not to derive a principle, we claim that there are consistent quantum gravity theories that obey it and possibly also those that do not, but rather to understand what its ramifications might be and how it can be tested.}  It is more likely that the principle or the formalism used is incorrect or inapplicable.  However the test of any principle is experiment.  If in our world, like in Mathur's, there are sufficiently large quantum corrections to quantum gravity to smear the singularity of a black hole to its very microscopic Schwarzschild radius, then S. Winitzki has pointed out that the same effect is likely to make a large difference, of order 10 percent, in the radius of a typical neutron star.  Models of neutron stars are improving, and it may not be long before such a discrepancy is observable.

One may also hope that the Mathurian smearing cures UV divergences.  For example, a photon whose energy is between the Planck scale and the Landau pole would be diffuse, and so its effective coupling constant may be lower than a naive renormalization group argument would predict.  This opens the possibility that the Landau pole is removed by a similar mechanism to the cancellation of UV divergences in string theory, even in the absence of a GUT unification of the gauge symmetries.

The elimination of horizons would solve the horizon problem, thus eliminating one motivation for inflation.  Another motivation for inflation, the scale invariant curvature fluctuations, may be explained by various noninflationary cosmological scenarios, such as \cite{BNV}.  The monopole formation problem appears to rely on the Higgs sector, which has never been measured.  In addition inflation solves the flatness problem, although the Einstein-Hilbert action dynamically generates irrelevant polynomials in the curvature which at early times (the UV) were very large and may have been extremized by a flat universe.  Thus one may hope that the above horizonless scenario compliments the scenario in Ref.~\cite{BNV} in the beginnings of the realization of an inflation-free cosmology.

\section* {Acknowledgement}

We would like to thank M.~Abou~Zeid, A.~Adams, D.~Anselmi, R.~Argurio, N.~Arkani-Hamed, A.~Barbieri, G.~Barnich, J.~D.~Bekenstein, N.~Bouatta, S.~Bolognesi, G.~Compere, T.~Evslin, M.~Fairbairn, F.~Ferrari, M.~Henneaux, A.~Keurentjes, S.~Kuperstein, C.~Maccaferri, L.~Martucci, P.~Menotti, P.~van Nieuwenhuizen, D.~Persson, S.~Rychkov, A.~Sen, P.~Silva, P.~Spindel, W.~Troost, A.~Vikman and S.~Winitzki for illuminating discussions.

The work of JE is partially supported by IISN - Belgium (convention 4.4505.86), by the ``Interuniversity Attraction Poles Programme -- Belgian Science Policy'' and by the European Commission RTN program HPRN-CT-00131, in which he is associated to K. U. Leuven.

\end{document}

\bibitem{}
,
{\it },
hep-th/.

\bibitem{}
,
{}
Nucl.\ Phys.\ B {\bf 266}, 709 (1986).

\bibitem{MH}
G. T. Horowitz and J. Maldacena,
{\it The black hole final state},
hep-th/0310281.

\bibitem{GP}
D. Gottesman and J. Preskill,
{\it Comment on ``The black hole final state''},
hep-th/0311269.

\bibitem{Ashtekar}
A. Ashtekar, M. Bojowald,
{\it Black hole evaporation: A paradigm},
hep-th/0504029.

\bibitem{StromingerLesHouches}
A. Strominger,
{\it Les Houches lectures on black holes},
hep-th/9501071.

\bibitem{Susskind}
J. Russo, L. Susskind and L. Thorlacius,
{\it The endpoint of Hawking evaporation},
hep-th/9206070.

\bibitem{ChristiansenFulling77}
S. M. Christiansen and S. A. Fulling,
{\it Trace anomalies and the Hawking effect},
Phys.Rev. {\bf D15} (1977) 2088.

\end{thebibliography}
\end{document}

\bibitem{dorey}
N.~Dorey,
{\it The BPS spectra of two-dimensional supersymmetric gauge theories with
twisted mass terms},
JHEP {\bf 9811} (1998) 005,
hep-th/9806056.

\bibitem{DHT}
N.~Dorey, T.~J.~Hollowood and D.~Tong,
{\it The BPS spectra of gauge theories in two and four dimensions},
JHEP {\bf 9905} (1999) 006,
hep-th/9902134.

\bibitem{HT}   A.~Hanany, D.~Tong,
{\it Vortices, instantons and branes},
hep-th/0306150.

\bibitem{SY-vortici}
M.~Shifman and A.~Yung,
{\it Non-abelian string junctions as confined monopoles},
hep-th/0403149.

\bibitem{HT2}
A.~Hanany and D.~Tong,
{\it Vortex strings and four-dimensional gauge dynamics},
JHEP {\bf 0404} (2004) 066,
hep-th/0403158.

\bibitem{nitta1}
Y.~Isozumi, M.~Nitta, K.~Ohashi and N.~Sakai,
{\it All exact solutions of a 1/4 Bogomol'nyi-Prasad-Sommerfield equation},
arXiv:hep-th/0405129.

\bibitem{nitta2}
M.~Eto, M.~Nitta and N.~Sakai,
{ \it Effective theory on non-Abelian vortices in six dimensions},
Nucl.\ Phys.\ B {\bf 701}, 247 (2004)
[arXiv:hep-th/0405161].

\bibitem{Shifman-yung1}
M.~Shifman and A.~Yung,
{\it Localization of non-Abelian gauge fields on domain walls at weak coupling
 (D-brane prototypes II)},
Phys.\ Rev.\ D {\bf 70} (2004) 025013,
hep-th/0312257.

\bibitem{Yung:2000uy}
A.~Yung,
{\it What do we learn about confinement from the Seiberg-Witten theory},
hep-th/0005088.

\bibitem{VY}
A.~I.~Vainshtein and A.~Yung,
{\it Type I superconductivity upon
monopole condensation in Seiberg-Witten  theory},
hep-th/0012250.

\bibitem{MY}
A.~Marshakov and A.~Yung,
{\it Non-Abelian confinement via Abelian flux tubes in softly broken N = 2  SUSY
QCD},
Nucl.\ Phys.\ B {\bf 647}  (2002) 3,
hep-th/0202172.

\bibitem{Shifman-yung2}
M.~Shifman and A.~Yung,
{\it Domain walls and flux tubes in N = 2 SQCD: D-brane prototypes},
Phys.\ Rev.\ D {\bf 67} (2003) 125007,
hep-th/0212293.

\bibitem{vortici}
R.~Auzzi, S.~Bolognesi, J.~Evslin, K.~Konishi and A.~Yung,
{\it Nonabelian superconductors: vortices and confinement in N = 2 SQCD},
Nucl.\ Phys.\ B {\bf 673} (2003) 187,
hep-th/0307287.

\bibitem{monovortice}
R.~Auzzi, S.~Bolognesi, J.~Evslin and K.~Konishi,
{\it Nonabelian monopoles and the vortices that confine them},
Nucl.\ Phys.\ B {\bf 686} 119 (2004),
hep-th/0312233.

\bibitem{GoddardNuytsOlive}
P.~Goddard, J.~Nuyts and D.~I.~Olive,
{\it Gauge Theories And Magnetic Charge},
Nucl.\ Phys.\ B {\bf 125} (1977) 1.

\bibitem{Weinberg}
E.~J.~Weinberg,
{\it Fundamental Monopoles In Theories With Arbitrary Symmetry Breaking},
Nucl.\ Phys.\ B {\bf 203} (1982) 445.

\bibitem{monopoli}
R.~Auzzi, S.~Bolognesi, J.~Evslin, K.~Konishi and H.~Murayama,
{\it Nonabelian monopoles},
hep-th/0405070.

\bibitem{tong-monopolo}
D.~Tong,
{\it Monopoles in the Higgs phase},
Phys.\ Rev.\ D {\bf 69} (2004) 065003,
hep-th/0307302.

\bibitem{Stefano}
S.~Bolognesi,
{\it The Holomorphic Tension of Vortices},
hep-th/0411075.

\bibitem{semi1}
J.~Preskill,
{\it Semilocal Defects},
Phys. Rev. D {\bf 46} (1992) 4218,
hep-ph/9206216.

\bibitem{semi2}
A.~Achucarro, A. C.~Davis, M.~Pickles and J.~Urrestilla,
{\it Fermion Zero Modes in $N=2$ Vortices},
Phys. Rev. D {\bf 68} (2003) 065006,
hep-th/0212125.

\bibitem{semi3}
K.~Evlampiev and A.~Yung,
{\it Flux Tubes on Higgs Branches in SUSY Gauge Theories},
Nucl. Phys. {\bf B662}  (2003) 230,
hep-th/0303047.

\bibitem{APS}
P. C.~Argyres, M. R.~Plesser and N.~Seiberg,
{\it The Moduli Space of N=2 SUSY QCD and Duality in N=1 SUSY QCD},
Nucl. Phys. {\bf B471}  (1996) 159,
hep-th/9603042.

\bibitem{CKM}
G.~Carlino, K.~Konishi and H.~Murayama,
{\it Dynamical Symmetry Breaking in Supersymmetric $SU(N_c)$ and $USp(2N_c)$ Gauge Theories},
Nucl. Phys. {\bf B590}  (2000) 37,
hep-th/0005076.


\bibitem{Bogomolny}
E.~B.~Bogomol'nyi,
{\it The stability of classical solutions},
Sov.\ J.\ Nucl.\ Phys.\ {\bf 24} (1976) 449.

\bibitem{AV}
A.~Achucarro and T.~Vachaspati,
{\it Semilocal and Electroweak Strings},
Phys.\ Rep.\  {\bf  327} (2000) 347-427,
hep-ph/9904229.

\bibitem{HSZ}
A.~Hanany, M.~J.~Strassler and A.~Zaffaroni,
{\it Confinement and strings in M{QCD}},
Nucl.\ Phys.\ B {\bf 513} (1998) 87
[arXiv:hep-th/9707244].

\bibitem{Hou}
Xin-rui Hou, {\it Abrikosov string in N=2 supersymmetric QED},
 Phys.\ Rev.\ D\ {\bf 63} (2001) 045015,
hep-th/0005119.

\bibitem{ken-rob}
R.~Auzzi and K.~Konishi,
{\it Non-universal corrections to the tension ratios in softly broken N = 2 SU(N) gauge theory,}
New J.\ Phys.\  {\bf 4} (2002) 59
[arXiv:hep-th/0205172].

\bibitem{prog}
M.~Shifman and A.~Yung,
{\it Non-Abelian Flux Tubes in SQCD: Supersizing World-Sheet Supersymmetry}
[arXiv:hep-th/0501211].

\bibitem{Gibbons9803203}
G.~W.~Gibbons,
{\it Branes as BIons},
Class.\ Quant.\ Grav.\  {\bf 16} (1999) 1471,
hep-th/9803203.

\bibitem{hanany-witten}
A.~Hanany and E.~Witten,
{\it Type IIB superstrings, BPS monopoles, and three-dimensional gauge
dynamics},
Nucl.\ Phys.\ B {\bf 492} (1997) 152,
hep-th/9611230.

\bibitem{witten-n=2}
E.~Witten,
{\it Solutions of four-dimensional field theories via M-theory},
Nucl.\ Phys.\ B {\bf 500} (1997) 3,
hep-th/9703166.

\bibitem{witten-n=1}
E.~Witten,
{\it Branes and the dynamics of {QCD}},
Nucl.\ Phys.\ B {\bf 507} (1997) 658,
hep-th/9706109.

\bibitem{Hori:1997ab}
K.~Hori, H.~Ooguri and Y.~Oz,
{\it Strong coupling dynamics of four-dimensional N = 1 gauge theories from  M
theory fivebrane},
Adv.\ Theor.\ Math.\ Phys.\  {\bf 1} (1998) 1,
hep-th/9706082.

\bibitem{deBoer:1997ap}
J.~de Boer and Y.~Oz,
{\it Monopole condensation and confining phase of N = 1 gauge theories via
M-theory fivebrane},
Nucl.\ Phys.\ B {\bf 511} (1998) 155,
hep-th/9708044.

\bibitem{Kneipp}
M.~Kneipp and P.~Brockill,
{\it  BPS string solutions in non-abelian Yang-Mills theories and confinement},
Phys.\  Rev.\ {\bf  D64} (2001) 125012,
hep-th/0104171.

\bibitem{Kneipp2}
M.~A.~C.~Kneipp,
{\it $Z_k$ string fluxes and monopole confinement in non-Abelian theories},
Phys.\ Rev.\ D {\bf 68} (2003) 045009,
hep-th/0211049.

\bibitem{Kneipp3}
M.~A.~C.~Kneipp,
{\it BPS $Z_k$ strings, string tensions and confinement in non-Abelian theories},
hep-th/0211146.

\bibitem{Kneipp4}
M.~A.~C.~Kneipp,
{\it Color superconductivity, $Z_N$ flux tubes and monopole confinement in
deformed N = 2* super Yang-Mills theories},
Phys.\ Rev.\ D {\bf 69} (2004) 045007,
hep-th/0308086.

\bibitem{Kneipp5}
M.~A.~C.~Kneipp,
{\it Color superconductivity, BPS strings and monopole confinement in ${\mathcal N}=2$ and ${\mathcal N}=4$ super Yang-Mills theories},
hep-th/0401234.

\bibitem{AchucaUrre}
A.~Achucarro and J.~Urrestilla,
{\it F-term strings in the Bogomolnyi limit are also BPS states},
JHEP {\bf 0408} (2004) 050,
hep-th/0407193.

\bibitem{Sen}
A.~Sen,
{\it Tachyon Condensation on the Brane Antibrane System},
JHEP {\bf 9808} (1998) 012,
hep-th/9805170.

\bibitem{MeMonodromy}
J.~Evslin,
{\it Twisted K-Theory from Monodromies},
JHEP {\bf 0305} (2003) 030,
hep-th/0302081.

\bibitem{KS}
I.~R.~Klebanov and M.~J.~Strassler,
{\it Supergravity and a Confining Gauge Theory: Duality Cascades and the $\chi$SB-Resolution of Naked Singularities},
JHEP {\bf 0008} (2000) 052,
hep-th/0007191.

\bibitem{MeMay}
J.~Evslin,
{\it The Cascade is a MMS Instanton},
hep-th/0405210.

\bibitem{Wati}
W.~Taylor,
{\it D2-branes in B fields},
hep-th/0004141.

\bibitem{semilocal1}
G.~W.~Gibbons, M.~E.~Ortiz, F.~Ruiz Ruiz and T.~M.~Samols,
{\it Semilocal strings and monopoles} ,
Nucl.\ Phys.\ B {\bf 385} (1992) 127,
hep-th/9203023.

\end{thebibliography}

\end{document}

I watched as my friend fell down the infinite throat of a black hole.  The Hawking radiation hit her very slowly, but her watch moved still more slowly and so she was hardly able to radiate any of it away.  Instead she got so hot that she almost looked like Hawking radiation herself, if I'd had an IR cutoff I wouldn't have been able to recognize her.  But I never took my eyes off of her and after a long time the black hole finished evaporating and I flew over to her and we compared notes.  My friend was always a finite distance away, always bigger than the Planck scale, and always in view until the end when the black hole was gone and it was safe for us to meet.  Therefore none of her information was lost, she could not even have passed the horizon.  And if she couldn't, then nobody can, and so the horizon can never grow.  Of course if I'd had worse instruments she would have disappeared, so my IR cutoff would yield some horizon size and would even destroy unitarity.  

I could always see her, so I know what she must have seen.  She scattered off of a lot of radiation near the horizon, but in the Schwarzschild geometry she would have passed right through.  So maybe in the quantum theory the matter is not concentrated at a singularity, but rather is smeared with a characteristic scale of order the Schwarzschild radius as in Mathur's fuzzball version of the D1-D5 system, with even a little bit of matter outside so that the mass inside is always just barely too low for a horizon to exist.  Then from her point of view she scattered off of this matter.  In fact she scattered a number of times, and so an insensitive observer may be led to believe that at the end she had been reduced to thermal radiation.  Below I will argue that conversely the apparently thermal Hawking radiation is not thermal, but rather consists of my old, multiply-scattered friends.

While I have neither sufficiently accurate instruments, nor sufficiently selfless friends nor a sufficiently long postdoc to perform the above experiment, the result relies upon a single hypothesis about black holes in an asymptotically Minkowski space.

\noindent
\textbf{Hypothesis} \textit{For any amount of time $T$, a stationary observer sufficiently far away from a given black hole will be able to see any object that he drops into the black hole for at least time $T$ and furthermore can treat that object classically.}

In particular this argument requires energy conservation.  Strictly speaking the energy is only well-defined for an observer at infinity and our observer stays a finite distance away, and so we will need to further assume that we can measure energies to any given accuracy by putting the observer far enough away and transporting him to and from this distance sufficiently long before and after the lifetime of the black hole.

In classical general relativity this hypothesis is the well known fact that an asymptotic observer can always see every object that has ever fallen into a black hole, although those objects will become very red.  In the Penrose diagram Fig. \ref{penclassico} one sees that at an arbitrary time $T$ the past lightcone of an observer outside of the black hole intersects all of the infalling matter.  

\begin{figure}[ht]
\begin{center}
\leavevmode
\epsfxsize 12  cm
\epsffile{penclassico.eps}
\end{center}
\caption{\footnotesize }
\label{penclassico}
\end{figure}

However the past lightcone never intersects the inside of the horizon, and so an observer outside of the horizon never has access to information about the inside of the horizon.  The observable region is depicted in the spacetime diagram Fig.~\ref{classico} where blue lines represent outward lightrays.

\begin{figure}[ht]
\begin{center}
\leavevmode
\epsfxsize 11  cm
\epsffile{classico.eps}
\end{center}
\caption{\footnotesize }
\label{classico}
\end{figure}

The Penrose diagram Fig.~\ref{penclassico} is extended to the other side of the horizon by assuming a classical evolution according to Einstein's equations, an approximation that is often assumed to be valid because the curvature near the horizon is small.  In this note however we will argue that this continuation is inconsistent with the inclusion of Hawking radiation, and so is apparently invalid in the quantum theory.  

If we include Hawking radiation these pictures change dramatically.  We will consider an asymptotically Minkowski spacetime, so that the black hole finishes evaporating at a finite time $T_1$ as measured by the distant observer.  The usual Penrose diagram associated with a radiating black hole is drawn in Fig.~\ref{penquantistico}

\begin{figure}[ht]
\begin{center}
\leavevmode
\epsfxsize 12  cm
\epsffile{penquantistico.eps}
\end{center}
\caption{\footnotesize }
\label{penquantistico}
\end{figure}
Here rather than each line corresponding to a piece of infalling matter, the lines represent the flow of energy.  Thus the energy crossing a surface is just the number of lines that intersect that surface.  This representation is useful because the fact that none of the lines ends enforces the conservation of energy, in particular the same number of lines enter and exit a compact region of spacetime and so the intersection number with a boundary must be zero.  As the lines only depict energy, they do not distinguish between infalling matter and Hawking radiation and so one may wish to include lines that continue through the event horizon representing the infalling matter plus lines with the opposite orientation exiting from the horizon that represent Hawking radiation.  However the lines here represent the total energy, and I will argue momentarily that the net energy flow into the horizon is zero and so no lines crossing the horizon are drawn in this representation.

Figure \ref{penquantistico} is a radial cross-section of a star collapsing to form a black hole.  For simplicity one may consider it to be spherically symmetric, although this is not essential.  Then each point in the Penrose diagram represents a sphere whose radius goes to zero on the line at the left hand side of the diagram, representing the center of mass of the configuration.  In particular given two times $T_0$ and $T$ as measured by a distant observer one may construct a compact surface with no boundaries by pasting together four components named $A,\ B,\ C,$\ and $D$.  $B$ is the trajectory of the observer from time $T_0$ to $T$ together with the entire sphere at the same radial coordinate at each moment in time, so for a stationary observer this is a cylinder.  $A$ and $C$ are the paths traced by lightrays that extend from the center of mass to the two boundary spheres of $B$ at times $T_0$ and $T$ respectively.  These are cones.  $D$ is just a line connecting the disks at the center of mass, it is one-dimensional as the sphere degenerates at the center of mass.  $D$ will not be important and is not even required for the surface to have no boundary.

As the union of $A,\ B,\ C,$\ and $D$ forms a compact boundary, the conservation of energy as measured by the distant observer, or at least the approximate conservation for an observer who is not quite at infinity, implies that the net energy flux through the union will be zero, or at least will be asymptotically zero as the observer is moved far away.  As $D$ is of a smaller dimension then the others in more than two dimensions and there is no singularity at or before time $T$, or more generally because $D$ follows the center of mass, no energy will flow through $D$.  If we consider $T_0$ to be sufficiently far in the past that negligible Hawking radiation has been emited from the configuration then the energy flow through $A$ will be arbitrarily close to the total mass $M$ of the collapsing star.  The energy $B(T)$ integrated over $B$ is equal to the Hawking radiation that has passed the observer by time $T$.  The energy $E(T)$ passing the surface $C$ is the energy that is still in the system at time $T$, that is the energy of the remaining infalling matter with a negative contribution from the Hawking radiation that has formed further from the horizon.  As the metric is well-behaved near $C$, one may perturb $C$ so that instead of having a null-direction it is a Cauchy surface on the interior of the radius of the distance observer.  The horizon is formed after this Cauchy surface, and so the energy entering the horizon cannot exceed $E(T)$ for any $T\leq T_1$.  Of course a real Cauchy surface should extend out to infinite radius, but the extra Hawking radiation passing a continuation of this surface to infinity is beyond the distant observer and is outbound.  This outbound energy never returns to the horizon and infact the horizon is not even in it's future lightcone, thus there is no need to extend the Cauchy surface beyond the radius of the observer.  

The total energy entering the union is $M$, the integral of energy over $A$.  The energy leaving the union consists of contributions $B(T)$, the radiation emited before time $T$, and $E(T)$, the amount of energy still in the system according to the distant observer at time $T$.  Thus the conservation of energy implies that
\begin{equation}
M=B(T)+E(T).
\end{equation}
Time $T_1$ is defined to be the moment at which the distant observer sees the black hole finish evaporating, and so as $T$ approaches $T_1$ the integral of the energy that has passed through $B$ approaches $M$.  Thus $B(T_1)=M$ and so
\begin{equation}
E(T_1)=0.
\end{equation}
This means that no energy enters the horizon, and so the horizon does not form and Fig. \ref{penquantistico} is unphysical.

For example a four-dimensional Schwarzschild solution has horizon radius
\begin{equation}
R=\frac{2GM}{c^2} \label{raggio}
\end{equation}
and so for any given horizon radius $R$, $c^2 R/2G$ units of mass must pass through the horizon.  However there will always exist a time $T<T_1$ such that the remaining mass of less than $c^2 R/2G$ on a Cauchy surface attached to that time.  As the horizon forms after this Cauchy surface, there is not enough energy around to form a horizon of radius $R$.  This is true for any $R$, and so no horizon forms.  

The above argument consists of a mix of selected facts from the classical and quantum descriptions, and is not the product of any consistent theory.  As such, it demonstrates nothing.  In particular the Hawking radiation only exists in the quantum theory, while I've been integrating energies of very red observers over slices of classical space.  Such integrals make no sense if the characteristic distances are so small that classical geometry no longer applies.  In particular, if my friend has shrunk to a size smaller than the Planck length, I would not be able to read her watch within the context of general relativity.

Of course the size of my friend, and thus the cutoff of the applicability of general relativity, depends on the Lorentz frame chosen.  In flat space if two observers with relative speeds near the speed of light both look at any object one is likely to see that object Lorentz contracted beyond the Planck scale and so one may claim that the classical approximation is invalid and one needs quantum gravity.   Here the infalling friend approaches the speed of light with respect to the distant observer, and so one may similarly fear that such a classical analysis is inapplicable.

For simplicity I will consider the case of a friend falling into a 4-dimensional Schwarzschild black hole, but again the generalization to other black holes, even those with infinite lifetimes, is straightforward.  The radius of the event horizon is again given by Eq.~(\ref{raggio}).  Although I have argued that really there is no event horizon, the strategy is to assume that there is a horizon and then arrive at a contradiction.  Imagine that my friend is a spherical shell distributed about the center of mass of the black hole with mass $m$.  Then after she falls in the horizon radius increases by $r$ units to
\begin{equation}
R^{\prime}=R+r=\frac{2G(M+m)}{c^2}.
\end{equation}

At every moment in the future I can see my friend, as my backward lightcone always intersects her.  She occupies a spherical shell with radius $r$.  $r$ is just a coordinate distance, to find the actual distance $d$ I need to integrate the metric $\sqrt{g_{rr}}$ over my past lightray from her feet to her head.  The metric is time dependent, but outside of the horizon at fixed radial coordinate $\rho$ with respect to the center of mass $g_{rr}$ is monotonically increasing with respect to time, because the difference between $\rho$ and the radius of the horizon is decreasing.  Thus a lower bound on the distance may be obtained by integrating the original Schwarzschild metric, that of a black hole with radius $R$.  Thus
\begin{eqnarray}
d&\geq&\int_{R}^{R+r}\sqrt{g_{rr}(\rho)}d\rho=\int_{R}^{R+r}\frac{1}{\sqrt{1-R/\rho}}d\rho\nonumber\\
&\cong&\int_{0}^{r}\sqrt{R/\rho^{\prime}}d\rho^{\prime}=2\sqrt{Rr}=\frac{4G\sqrt{Mm}}{c^2}
\end{eqnarray}
where in the second line we dropped terms surpressed by factors of order $r/R$. In particular if one considers my friend to be $n$ smaller friends each of mass $m/n$ then one finds a lower bound on the asymptotic thickness of all $n$ little friends together which is greater by a factor of $\sqrt{n}$.  As $n$ is arbitrary, the thickness of my friend rather than asymptoting to zero or to the Planck scale appears to explode.  This is not so surprising, as the distant observer also sees the black hole's throat monotically growing until near the end of the hole's lifetime, and the falling matter is presumably distributed throughout this lengthening throat.  

The assumption that the initial and final geometry is Schwarzschild is very strong.  More generally the event horizon could be growing with some velocity $v$.  In this case instead of $g_{rr}$ tending to infinity, the fact that light near the horizon travels parallel to the horizon implies instead that
\begin{equation}
\frac{g_{tt}}{g_{rr}}=v^2
\end{equation}
and so $g_{rr}$ tends to a value much larger than the Planck scale if $g_{tt}$ also tends to a value much larger than the Planck scale.  While I cannot prove that this is in general the case, in the limiting case $v=c$ the spacetime near the horizon is Minkowski and so $g_{rr}=1$, this is because matter outside of the horizon does not affect $v$ in a spherically symmetric configuration, and matter inside the horizon would bend the light towards it and so make $v<c$.  The other limiting case, $v=0$, we have just seen leads to an infinite distance $d$.  Thus one may expect that for a general $v$ the metric component $g_rr$ will be greater than one, and in particular well above the Planck scale.  Thus the distance over which my friend friend is smeared will be well above the Planck scale.

While this certainly does not prove that the classical approximation is justified, it implies that the approximation does at least pass one consistency check.  It would be nice to have an explicit metric for collapsing matter and check that the linear density does not shrink to the Planck scale.  If you have one, please send it to me.

General relativity tells us that an infalling friend passes through the horizon without even noticing it.  However this appears to be inconsistent with the observations of the distant observer, who observed that a Cauchy surface before the horizon contains no mass and so no horizon may form.  In fact this argument is a bit too fast.  If the area bounded by the horizon contains no net mass then an analogy to the Schwarzschild black hole may lead us to believe that the horizon has no area, in other words the distant observer has seen the mouth of the throat close.  It does not however determine whether the throat itself had zero volume at the time the black hole finished evaporating, nor whether my friend was still inside.  The conservation of energy does imply however that the total energy left in the throat at the moment when it pinches off from the rest of spacetime is zero, that is the positive energy of the infalling matter precisely cancels the negative energy of the infalling part of the Hawking radiation.  After the pinch of course the throat no longer has an asymptotic region and so there is not any notion of asymptotic energy.  However if the throat volume has gone to zero before it pinched off then we may hope that the infalling matter and antimatter have annihilated and so, as in the Maldacena-Horowitz case, entanglement has preserved all of the information.  In other words, colliding pairs of particles have opposite quantum numbers, as did the created pairs of Hawking radiation, and so the Hawking radiation is a clone of my friend once my friend has been destroyed.  Equivalently one may think that my friend travelling forward in time, then backward for awhile as infalling Hawking radiation, and then forward again.  This is just a more complicated description of Figure \ref{penquantistico}.  In particular it leads us to consider the possibility that the Hawking radiation is just the infalling matter after a series of scatterings, as in Maldacena-Horowitz and Gottesman-Preskill.  This is consistent with the fact that there is always just as much radiation as there is infalling matter.

The other possibility is that the throat still had a positive volume when it pinched off, in which case we have described one universe pinching into two.  While such a process is obviously beyond the naive treatment of this article, I am unable to refute the claim that my friends part of the universe could split off without him noticing it.  However whether the throat's length is finite or infinite when the mouth closes, in neither case does the region bounded by the horizon have a positive energy as measured from asymptopia before the mouth closed.  In particular, if the horizon area vanishes when the amount of energy inside vanishes, then any horizon must have zero area.  

The distant observer saw his friend's watch almost stop as she was bombarded with radiation.  In fact, because the friend is a spherical shell, Hawking radiation of energy equal to the total infalling mass of the black hole blasted its way through her before time $T_1$ according to the distant observer, that is before she reached the horizon.  Thus the distant observer concludes that his friend was blasted with radiation during a time that for her was so short that she could not have radiated it away herself.  In short, the distant observer concludes that the friend not only would be aware of the horizon but as she approaches it, it must blow up in her face.  More precisely, the friend, like the observer, can always see everything that has ever fallen into the black hole distributed throughout a long throat.  Gradually Hawking radiation leaks out of this throat.  As the friend appraoches, the black hole begins to age very quickly and the throat Lorentz contracts, so the radiation becomes ever more intense.  Once the friend is almost at the horizon the entire black hole evaporates away in what for her is a very short amount of time.  She scatters off of this massive burst of radiation and in the process is vaporized into almost thermal radiation herself, in particular an insensitive observer might think that she had become Hawking radiation.  

However this scattering happens outside of any horizon and no information is lost.  Notice that if the distant observer applies a Bogoliubov transformation to his friend to determine what she sees, he concludes that she observes no radiation. However if we are to interpret the Hawking radiation as being made of scattered infalling friends, then the friends must interact in order to scattered.  Perhaps the Bogoliubov transformation does not capture the interactions seen by the friend because these interactions are with other matter, friends that have already fallen in, and not with the vacuum.


Thus the entire black hole collapse and evaporation may just be a scattering process, in which material shrinks towards a minimum radius, in Mathur's D1-D5 fuzzball system this is seen to be a minimal distance compatible with the uncertainty principle, and then after a series of scatterings it escapes to infinity.  An outside observer sees this as a bounce that occurs over a period of $10^{70}$ years involving particles moving near the speed of light into and out of a very long throat, an infalling friend sees much smaller distances and times.  In short, the above diagrams appear to fail before the event horizon, and thus far from the formation of a singularity.  A more consistent spacetime diagram is drawn in Figure \ref{quantistico}.  

\begin{figure}[ht]
\begin{center}
\leavevmode
\epsfxsize 10  cm
\epsffile{quantistico.eps}
\end{center}
\caption{\footnotesize }
\label{quantistico}
\end{figure}

A real observer will have instruments with a finite sensitivity.  As the friend falls in she blushes until she is very far into the infrared.  However she never gets infinitely far into the infrared, in fact there is a maximal redshift that is finite before she has finished being radiated away.  If the sensitivity of the observers instruments is less than the energy scale of this maximal redshift then it will appear has though she has disappeared.  A sensitivity limit is equivalent to an IR cutoff, and so in the IR cutoff theory there will be a horizon, and its position will depend on the cutoff.  Now the nonunitary of the theory with a horizon may just be interpreted as the usual nonunitary in a theory with an IR cutoff.  It is harmless, as it goes away as the IR cutoff is taken to zero, in fact as the IR cutoff is reduced to the finite value equal to the maximal redshift.

\end{document}

This note examines the implications of the hypothesis that this fact remains true in the quantum theory, more precisely, in the case in which black holes evaporate out of existence after a finite period of time.

I will argue that because the observer can always observe his friend's watch from the moment that they separate to the moment that they reunite, and it changes continuously, the friend has never passed the event horizon.  Similarly, nothing may pass the event horizon during the finite lifetime of the black hole, and so the event horizon cannot grow.  

This is a conjecture about the observations of a distant observer.  However as the observer is always able to watch his friend who is falling into the black hole, the observervations of the distant observer are sufficient to deduce those of the infalling friend.  In particular the friend's observations are inconsistent with the continuation of the Schwarzschild geometry inside of the event horizon.  For example, in the Schwarzschild geometry the friend should not even be aware that she is passing the horizon.  However the observations of the distant observer imply that, on the contrary, as the friend approaches the black hole ages very quickly and explodes altogether before she can reach the horizon.   Equivalently as she approaches all of the radiation from the long throat is Lorenz contracted into a finite length, and she scatters off of all of the energy of the black hole.

This does appear to be consistent with event horizons that are filled with matter, as in the D1-D5 system [Mathur].  In fact it may be that not quite enough matter can fit in the horizon for the horizon to exist, and so no horizon exists in the quantum theory.  This is even stronger than the above claim that horizons do not grow.

The situation is very different if the distant observer's instruments have a finite sensitivity, which is equivalent to introducing an IR cutoff.  In this case, if the cutoff is sufficiently high, then the friend may become too red to be observed.  At this point the observer would conclude that she has fallen through the horizon, and her information has been lost.  Thus one would expect that size of the horizon to depend on the IR cutoff.  As is typical when one introduces a cutoff, one expects the corresponding information loss to destroy the unitary of the cutoff theory.  However as the cutoff is taken to zero the horizon shrinks away and unitarity is restored.

An interesting test of this proposal would be to calculate the amount of matter inside of each radius in the D1-D5 system and to show that it is always just barely too small for there to be an event horizon at this radius.  In Ref.~[Mathur] it was shown that to leading order the radius of a D1-D5 system is of order the Schwarzschild radius.  Thus the above proposal is to reproduce the known coefficient and then to show that the leading order correction to the radius over which the matter is smeared is positive, and so the matter is too smeared to form an event horizon.  More precisely one needs to check that there is insufficient matter to form a black hole inside of every radius.

It would be interesting to see whether horizons are also forbidden.


\begin{thebibliography}{23}

\bibitem{Sagnotti1}
M. H. Goroff and A. Sagnotti,
{\it Quantum Gravity at Two Loops},
Phys.\ Lett.\ B {\bf 160}, 81 (1985).

\bibitem{Sagnotti2}
M. H. Goroff and A. Sagnotti,
{\it The Ultraviolet Behavior of Einstein Gravity},
Nucl.\ Phys.\ B {\bf 266}, 709 (1986).

\bibitem{WittenCoset}
E. Witten,
{\it On string theory and black holes},
Phys.Rev. {\bf D44} (1991) 314-324.

\bibitem{CGHS}
C. Callan, S. Giddings, J. Harvey and A. Strominger,
{\it Evanescent black holes},
hep-th/9111056.

\bibitem{Lunin1}
O. Lunin and S. D. Mathur,
{\it AdS/CFT duality and the black hole information paradox},
hep-th/0109154.


\bibitem{Lunin2}
O. Lunin, J. Maldacena and L. Maoz
{\it Gravity solutions for the D1-D5 system with angular momentum},
hep-th/0212210.

\bibitem{Mathur1}
S. D. Mathur, A. Saxena and Y. K. Srivastava,
{\it Constructing ``hair'' for the three charge hole},
hep-th/0311092.

\bibitem{Mathur2}
S. D. Mathur,
{\it What are the states of a black hole?},
hep-th/0401115.

\bibitem{Lunin3}
O. Lunin,
{\it Adding momentum to D1-D5 system},
hep-th/0404006.

\bibitem{Mathur3}
S. Giusto, S. D. Mathur and A. Saxena,
{\it Dual geometries for a set of 3-charge microstates},
hep-th/0405017.

\bibitem{Mathur4}
S. Giusto, S. D. Mathur and A.~Saxena,
{\it 3-charge geometries and their CFT duals},
hep-th/0406103.

\bibitem{Mathur5}
S. Giusto, S. D. Mathur,
{\it Geometry of D1-D5-P bound states},
hep-th/0409067.

\bibitem{Mathur6}
S. Giusto, S. D. Mathur,
{\it Fuzzball geometries and higher derivative corrections for extremal holes},
hep-th/0412133.

\bibitem{Mathur7}
S. D. Mathur,
{\it The fuzzball proposal for black holes: an elementary review},
hep-th/0502050.

\bibitem{thooft85}
G. 't Hooft,
{\it The Quantum Structure of a Black Hole},
Nucl.\ Phys.\ B {\bf 256}, 727 (1985).

\bibitem{thooft94}
G. 't Hooft, C. R. Stephens and B. F. Whiting,
{\it Black Hole Evaporation Without Information Loss},
Class.\ Quantum Grav.\ {\bf 11}, 621 (1994).

\bibitem{thooft96}
G. 't Hooft,
{\it The Scattering Matrix Approach for the Quantum Black Hole: An Overview},
gr-qc/9607022.

\bibitem{thooft05}
G. 't Hooft,
{\it The Holographic Mapping of the Standard Model onto the Black Hole Horizon.  Part I. Abelian Vector Field, Scalar Field and BEH Mechanism},
gr-qc/0504120.

\bibitem{EP1}
A. E. Faraggi and M. Matone,
{\it Equivalence Principle, Planck Length and Quantum Hamilton-Jacobi Equation},
hep-th/9809125.

\bibitem{EP2}
A. E. Faraggi and M. Matone,
{\it Equivalence Principle: Tunneling, Quantized Spectra and Trajectories from the Quantum HJ Equation},
hep-th/9809126.

\bibitem{EP3}
A. E. Faraggi and M. Matone,
{\it The Equivalence Postulate of Quantum Mechanics},
hep-th/9809127.

\bibitem{ChristiansenFulling77}
S. M. Christiansen and S. A. Fulling,
{\it Trace anomalies and the Hawking effect},
Phys.Rev. {\bf D15} 2088 (1977).

\bibitem{Unruh}
W. G. Unruh,
{\it Notes on Black Hole Evaporation},
Phys.Rev. {\bf D14} 870 (1976).

\bibitem{Strom}
A. Strominger,
{\it Fadeev-Popov Ghosts and 1+1 Dimensional Black Hole Evaporation},
hep-th/9205028.

\bibitem{Horava}
E. G. Gimon and P.~Ho{\v r}ava,
{\it Over-Rotating Black Holes, Godel Holography and the Hypertube},
hep-th/0405019.

\bibitem{BNV}
A. Nayeri, R. H. Brandenberger, C. Vafa,
{\it Producing a Scale-Invariant Spectrum of Perturbations in a Hagedorn Phase of String Cosmology},
hep-th/0511140.

\end{thebibliography}
\end{document}